\documentclass[a4paper,11pt]{article}
\pdfoutput=1 

\usepackage{jheppub} 

\usepackage[T1]{fontenc} 

\title{\boldmath Remnants, fermions' tunnelling and effects of quantum gravity}

\author[a]{D.Y. Chen,}
\author[a]{Q.Q. Jiang,}
\author[b]{P. Wang}
\author[b]{and H. Yang}

\affiliation[a]{Institute of Theoretical Physics, \\China West Normal University,\\Nanchong, 637009, China}
\affiliation[b]{Center for Theoretical Physics, \\College of Physical Science and Technology, \\Sichuan University,\\Chengdu, 610051, China}

\emailAdd{dchen@cwnu.edu.cn}
\emailAdd{qqjiang@cwnu.edu.cn}
\emailAdd{pengw@scu.edu.cn}
\emailAdd{hyanga@scu.edu.cn}

\abstract{The remnants are investigated by fermions' tunnelling from a 4-dimensional charged dilatonic black hole and a 5-dimensional black string. Based on the generalized uncertainty principle, effects of quantum gravity are taken into account. The quantum numbers of the emitted fermions affects the Hawking temperatures. For the black hole, the quantum gravity correction slows down the increase of the temperature, which leads to the remnant left in the evaporation. For the black string, the existence of the remnant is determined by the quantum gravity correction and effects of the extra compact dimension.}

\begin{document}
\maketitle
\flushbottom

\section{Introduction}

The standard Hawking formula predicts the complete evaporation of black holes. In the original research \cite{SWH}, the formula was gotten in the frame of quantum field theory on curved spacetime. It is based on the Heisenberg uncertainty principle (HUP). Therefore, it is natural to find that the complete evaporation is a direct consequence of the HUP. The semi-classical tunnelling method put forward by Parikh and Wilczek is an effective way to research on Hawking radiation \cite{PW}. With the consideration of the variable background spacetime, the tunnelling behavior of photons across the horizons was described veritably. The corrected temperatures were gotten and higher than the standard Hawking temperatures \cite{SWH}. This result indicates that the variable spacetimes speed up the increases of the temperatures and the black holes evaporate completely. The extension of this method to the tunnelling radiation of massive scalar particles was found in the subsequent work \cite{ZZ,JWC}. The Hamilton-Jacobi ansatz is another version of the tunnelling method \cite{ANVZ,KM1}. Adopting this ansatz, the standard Hawking temperatures were recovered by fermions' tunnelling across the horizons of the black holes \cite{KM2}. All of these results lead to that black holes evaporate completely and there are no remnants left \cite{AAS}.

On the other hand, various theories of quantum gravity, such as string theory, loop quantum gravity and quantum geometry, predict the existence of the minimal observable length \cite{PKT,ACV,KPP,LJG,GAC,NIC2}. This view is supported by the Gedanken experiment \cite{FS}. An effective model to realize this minimal length is the generalized uncertainty principle (GUP),

\begin{eqnarray}
\Delta x \Delta p \geq \frac{\hbar}{2}\left[1+ \beta (\Delta p)^2\right],
\label{eq1.1}
\end{eqnarray}

\noindent which is derived by the modified fundamental commutation relations. $\beta = \beta_0 \frac{l^2_p}{\hbar^2}$ is a small value, $\beta_0 <10^{34}$ is a dimensionless parameter and $l_p$ is the Planck length. Kempf et. al. first modified the commutation relations and got $\left[x_i,p_j\right]= i \hbar \delta_{ij}\left[1+ \beta p^2\right]$, where $x_i$ and $p_i$ are position and momentum operators defined respectively as  \cite{KMM}

\begin{eqnarray}
x_i &=& x_{0i}, \nonumber\\
p_i &=& p_{0i} (1 + \beta p^2),
\label{eq1.2}
\end{eqnarray}

\noindent $x_{0i}$ and $p_{0j}$ satisfy the canonical commutation relations $\left[x_{0i},p_{0j}\right]= i \hbar \delta_{ij}$. The modification is not unique. Other modifications are referred to \cite{AK,FB,ADV,DV,NIC3}.

These modifications play an important role on the black hole physics. Based on the GUP, it was found that there is no existence of black holes at LHC in \cite{AFA}. The black hole thermodynamics was discussed in \cite{XW,BJM,KP,ZDM}, respectively. The relation between the area and entropy and the corrected Hawking temperatures were gotten. An interested result is that the remnants exist in black holes' evaporation \cite{ACS,SGC,BG,LX,NS,NIC1}. Incorporating the GUP into the tunnelling radiation in scalar fields, the corrected Hawking temperatures in the Schwarzschild and the noncommutative spacetimes were obtained \cite{NS,NM}. Using the modified commutation relation between the radial coordinate and the conjugate momentum and considering the natural cutoffs as minimal and maximal momentum, the tunnelling rates were derived in \cite{NS}. The interesting result is that the minimal mass and the maximum temperature in the scalar field were found.

In this paper, taking into account effects of quantum gravity, we investigate the tunnelling radiation of fermions from a 4-dimensional charged dilatonic black hole and a 5-dimensional black string. The remnants are discussed by the corrected Hawking temperatures. The temperatures are affected by the quantum numbers (mass, charge and energy) of the emitted fermions. For the dilatonic black hole, the quantum gravity correction slows down the increase of the Hawking temperature. It is natural to lead to the remnant existed in the evaporation. In the black string spacetime, the quantum gravity correction and the effect of the extra compact dimension affect the evaporation.

The rest is outlined as follows. In the next section, based on the modified commutation relations put forward in \cite{KMM}, we modify the Dirac equation in curved spacetime. In section 3, with the consideration of effects of quantum gravity, the fermion's tunnelling from the charged dilatonic black hole is investigated and the remnant is derived. In section 4, we investigate the fermion's tunnelling from the black string. The evaporation of the string is discussed. Section 5 is devoted to our conclusion.

\section{Generalized Dirac equation}

In this section, we adopt the modified operators of position and
momentum in eqn. (\ref{eq1.2}) to modify  the Dirac equation in
curved spacetime. To achieve this purpose, we first introduce the
GUP into Dirac equation. We then generalize Dirac equation to
curved background by standard process. Respecting covariance is
certainly of importance during the derivations. Under this
constraint, the modification of Dirac equation in flat spacetime
based on GUP can be uniquely determined  \cite{NK,HBH,OS,MK}. The
square of momentum operators is

\begin{eqnarray}
p^2 &=& p_i p^i = -\hbar^2 \left[ {1 - \beta \hbar^2 \left( {\partial _j \partial ^j} \right)}
\right]\partial _i \cdot \left[ {1 - \beta \hbar^2 \left( {\partial ^j\partial _j } \right)}
\right]\partial ^i\nonumber \\
&\simeq & - \hbar ^2\left[ {\partial _i \partial ^i - 2\beta \hbar ^2
\left( {\partial ^j\partial _j } \right)\left( {\partial
^i\partial _i } \right)} \right].
\label{eq2.1}
\end{eqnarray}

\noindent In the last step, the higher order terms of $\beta $ are neglected. In the theory of quantum gravity, the generalized frequency takes on the form as \cite{WG}

\begin{eqnarray}
\tilde \omega = E( 1 - \beta E^2),
\label{eq2.2}
\end{eqnarray}

\noindent with the definition of energy operator $ E = i \partial _t $. Using the energy mass shell condition $ p^2 + m^2 = E^2 $, we get the generalized expression of energy \cite{NS,WG,NK,HBH}

\begin{eqnarray}
\tilde E = E[ 1 - \beta (p^2 + m^2)].
\label{eq2.3}
\end{eqnarray}

Then, in flat background, the modified  Dirac equation based on GUP follows straightforwardly as in  \cite{NK}. In curved spacetime, the Dirac equation with an electromagnetic field takes on the form as

\begin{eqnarray}
i\gamma^{\mu}\left(\partial_{\mu}+\Omega_{\mu}+\frac{i}{\hbar}eA_{\mu}\right)\psi+\frac{m}{\hbar}\psi=0,
\label{eq2.4}
\end{eqnarray}

\noindent where $\Omega _\mu \equiv\frac{i}{2}\omega _\mu\, ^{a b} \Sigma_{ab}$, $\Sigma_{ab}= \frac{i}{4}\left[ {\gamma ^a ,\gamma^b} \right]$, $\{\gamma ^a ,\gamma^b\}= 2\eta^{ab}$, $\omega _\mu\, ^{ab}$ is the spin connection defined by  $\omega_\mu\,^a\,_b=e_\nu\,^a e^\lambda\,_b \Gamma^\nu_{\mu\lambda} -e^\lambda\,_b\partial_\mu e_\lambda\,^a$, $\Gamma^\nu_{\mu\lambda}$ is the ordinary connection and $e^\lambda\,_b$ is the tetrad. The Greek indices are raised and lowered by the curved metric $g_{\mu\nu}$, while the Latin indices are governed by the flat metric $\eta_{ab}$. The construction of a tetrad satisfies the following relations

\begin{equation}
g_{\mu\nu}= e_\mu\,^a e_\nu\,^b \eta_{ab},\hspace{5mm} \eta_{ab}=g_{\mu\nu} e^\mu\,_a e^\nu\,_b, \hspace{5mm} e^\mu\,_a e_\nu\,^a=\delta^\mu_\nu, \hspace{5mm} e^\mu\,_a e_\mu\,^b = \delta_a^b.
\label{eq2.5}
\end{equation}

\noindent Therefore, it is readily to construct the $\gamma^\mu$'s in curved spacetime as

\begin{equation}
\gamma^\mu = e^\mu\,_a \gamma^a, \hspace{7mm} \left\{ {\gamma ^\mu,\gamma ^\nu } \right\} = 2g^{\mu \nu }.
\label{eq2.6}
\end{equation}

\noindent To modify the Dirac equation, we rewrite eqn. (\ref{eq2.4}) as

\begin{eqnarray}
-i\gamma^{0}\partial_{0}\psi=\left(i\gamma^{i}\partial_{i}+i\gamma^{\mu}\Omega_{\mu}+i\gamma^{\mu}
\frac{i}{\hbar}eA_{\mu}+\frac{m}{\hbar}\right)\psi,
\label{eq2.7}
\end{eqnarray}

\noindent namely,

\begin{eqnarray}
i\partial_{0}\psi= -\gamma_{0}\left(i\gamma^{i}\partial_{i}+i\gamma^{\mu}\Omega_{\mu}+i\gamma^{\mu}
\frac{i}{\hbar}eA_{\mu}+\frac{m}{\hbar}\right)\psi.
\end{eqnarray}

\noindent The left hand side of the above equation is related to the energy. Using the generalized expression of energy (\ref{eq2.3}), we get the modified Dirac equation as follows \cite{NK,HBH}

\begin{eqnarray}
i\partial_{0}\Psi &=& -\gamma_{0}\left(i\gamma^{i}\partial_{i}+i\gamma^{\mu}\Omega_{\mu}+i\gamma^{\mu}
\frac{i}{\hbar}eA_{\mu}+\frac{m}{\hbar}\right)\left(1-\beta p^2-\beta m^{2}\right)\Psi \nonumber \\
&=& -\gamma_{0}\left(i\gamma^{i}\partial_{i}+i\gamma^{\mu}\Omega_{\mu}+i\gamma^{\mu}
\frac{i}{\hbar}eA_{\mu}+\frac{m}{\hbar}\right)\left(1+\beta\hbar^{2}\partial_{j}\partial^{j}-\beta m^{2}\right)\Psi.
\end{eqnarray}

\noindent The last equal sign was derived by the expression of the square of momentum operators in eqn. (\ref{eq2.1}) and the neglect of the higher order terms of $\beta$. In this equation, $\Psi$ is the generalized Dirac field. Thus the modified Dirac equation in curved spacetime is

\begin{eqnarray}
-i\gamma^{0}\partial_{0}\Psi=\left(i\gamma^{i}\partial_{i}+i\gamma^{\mu}\Omega_{\mu}+i\gamma^{\mu}
\frac{i}{\hbar}eA_{\mu}+\frac{m}{\hbar}\right)\left(1+\beta\hbar^{2}\partial_{j}\partial^{j}-\beta m^{2}\right)\Psi,
\label{eq2.8}
\end{eqnarray}

\noindent which is rewritten as

\begin{eqnarray}
\left[i\gamma^{0}\partial_{0}+i\gamma^{i}\partial_{i}\left(1-\beta m^{2}\right)+i\gamma^{i}\beta\hbar^{2}\left(\partial_{j}\partial^{j}\right)\partial_{i}+\frac{m}{\hbar}
\left(1+\beta\hbar^{2}\partial_{j}\partial^{j}-\beta m^{2}\right)\right.\nonumber \\
\left.+i\gamma^{\mu}\frac{i}{\hbar}eA_{\mu}\left(1+\beta\hbar^{2}\partial_{j}\partial^{j}-\beta m^{2}\right)+i\gamma^{\mu}\Omega_{\mu}\left(1+\beta\hbar^{2}\partial_{j}\partial^{j}-\beta m^{2}\right)\right]\Psi= 0.
\label{eq2.9}
\end{eqnarray}

\noindent When $eA_{\mu} = 0$, it describes the Dirac equation without the electromagnetic field. In the following sections, we adopt eqn. (\ref{eq2.9}) to investigate fermions' tunnelling across the horizons of the 4-dimensional and the 5-dimensional spacetimes. If one only considers the modification of momenta in the Dirac equation. It is not covariant. From covariance, therefore, we modified the momenta and energy in the Dirac equation.

\section{The remnant in the 4-dimensional dilatonic black hole}

The general solution of dilatonic black holes \cite{GHSHH} was derived from the action

\begin{eqnarray}
S = \int {dx^4 \sqrt{-g} \left[-R + 2(\Delta \Phi)^2 + e^{-2\alpha \Phi}F^2 \right] },
\label{eq3.1}
\end{eqnarray}

\noindent which describes the standard matter, gravity coupled to a Maxwell field and a dilaton.  $\alpha$  is a parameter expressed the strength of coupling of the dilation field $\Phi$ to the Maxwell field $F$. It reduces to the usual Einstein-Maxwell scalar theory when $\alpha = 0$, while it is part of the low energy action of string theory when $\alpha = 1$. The metric of the spherically symmetric charged dilatonic black hole is given by

\begin{eqnarray}
ds^{2}=-f\left(r\right)dt^{2}+ \frac{1}{f\left(r\right)}dr^{2}+R^{2}(r)\left(d\theta^{2}+\sin^{2}\theta d\phi^{2}\right),
\label{eq3.2}
\end{eqnarray}

\noindent with the electromagnetic potential $A_{\mu}= \left(A_{t},0,0,0\right)= \left(\frac{Q}{r},0,0,0\right)$, where

\begin{eqnarray}
R\left(r\right) &=& r\left(1-\frac{r_-}{r}\right)^{\frac{\alpha^2}{1+\alpha^2}}, \nonumber\\f\left(r\right) &=& \left(1-\frac{r_+}{r}\right)\left(1-\frac{r_-}{r}\right)^{\frac{1-\alpha^2}{1+\alpha^2}} .
\label{eq3.3}
\end{eqnarray}

\noindent The event horizon is located at $r = r_+$ for all $\alpha$ and $r_+>r_-$. The mass and charge of the black hole are represented by $ M = \frac{r_+}{2} + \frac{r_-}{2}\left(\frac{1-\alpha^2}{1+\alpha^2}\right)$ and  $Q = \sqrt{\frac{r_+r_-}{1+\alpha^2}}$, respectively.

For a spin-1/2 fermion, there are two states corresponding to spin up and spin down. In this paper, we only consider the state with spin up without loss of generality. The investigation of the state with spin down is parallel. The motion of a fermion in the dilaton black hole obeys the generalized Dirac equation (\ref{eq2.9}). To describe the motion, we first suppose that the wave function of the fermion with spin up state takes on the form as

\begin{eqnarray}
\Psi=\left(\begin{array}{c}
A\\
0\\
B\\
0
\end{array}\right)\exp\left(\frac{i}{\hbar}I\left(t,r,\theta,\phi\right)\right),
\label{eq3.4}
\end{eqnarray}

\noindent where $I$ is the action of the fermion with spin up state and $A$ and $B$ are functions of $(t , r , \theta , \phi)$. To solve the equation (\ref{eq2.9}), one should construct gamma matrices. The construction of gamma matrices is relied on a tetrad. It is straightforward to get the tetrad from the metric (\ref{eq3.2}) as

\begin{eqnarray}
e_{\mu}^a = \rm{diag}\left(\sqrt{f}, 1/\sqrt{f}, R, R\sin\theta \right).
\label{eq3.5}
\end{eqnarray}

\noindent Then the gamma matrices is constructed as

\begin{eqnarray}
\gamma^{t}=\frac{1}{\sqrt{f\left(r\right)}}\left(\begin{array}{cc}
0 & \rm I\\
-\rm I & 0
\end{array}\right), &  & \gamma^{\theta}=\sqrt{g^{\theta\theta}}\left(\begin{array}{cc}
0 & \sigma^{1}\\
\sigma^{1} & 0
\end{array}\right),\nonumber \\
\gamma^{r}=\sqrt{f\left(r\right)}\left(\begin{array}{cc}
0 & \sigma^{3}\\
\sigma^{3} & 0
\end{array}\right), &  & \gamma^{\phi}=\sqrt{g^{\phi\phi}}\left(\begin{array}{cc}
0 & \sigma^{2}\\
\sigma^{2} & 0
\end{array}\right).
\label{eq3.6}
\end{eqnarray}

\noindent In the above equations, $\sqrt{g^{\theta\theta}} = R^{-1}$, $\sqrt{g^{\phi\phi}} =(R\sin\theta)^{-1}$, $\rm I$ is the unit matrix, $\sigma ^i$ are the Pauli matrices,

\begin{eqnarray}
\sigma^{1}=\left(\begin{array}{cc}
0 & 1\\
1 & 0
\end{array}\right), \quad \sigma ^{2}=\left(\begin{array}{cc}
0 & -i\\
i & 0
\end{array}\right), \quad \sigma ^{3}=\left(\begin{array}{cc}
1 & 0\\
0 & -1
\end{array}\right).
\label{eq3.7}
\end{eqnarray}

\noindent We insert the gamma matrices and the wave function into the equation (\ref{eq2.9}) and divide by the exponential term. Applying the WKB approximation, we get the resulting equations to leading order in $\hbar$. They are decoupled into four equations

\begin{eqnarray}
-B\frac{1}{\sqrt{f}}\partial_{t}I-B\left(1-\beta m^{2}\right)\sqrt{f}\partial_{r}I - Am\beta\left[g^{rr}\left(\partial_{r}I\right)^{2}+g^{\theta\theta}\left(\partial_{\theta}I\right)^{2}+
g^{\phi\phi}\left(\partial_{\phi}I\right)^{2}\right]\nonumber\\
+B\beta\sqrt{f}\partial_{r}I \left[g^{rr}\left(\partial_{r}I\right)^{2}+g^{\theta\theta}\left(\partial_{\theta}I\right)^{2}
+g^{\phi\phi}\left(\partial_{\phi}I\right)^{2}\right]+Am\left(1-\beta m^{2}\right)\nonumber\\
-B\frac{eA_t}{\sqrt f}\left[1-\beta m^{2} -\beta\left(g^{rr}\left(\partial_{r}I\right)^{2}+g^{\theta\theta}\left(\partial_{\theta}I\right)^{2}+
g^{\phi\phi}\left(\partial_{\phi}I\right)^{2}\right)\right]
= 0,
\label{eq3.8}
\end{eqnarray}

\begin{eqnarray}
A\frac{1}{\sqrt{f}}\partial_{t}I-A\left(1-\beta m^{2}\right)\sqrt{f}\partial_{r}I-Bm\beta\left[g^{rr}\left(\partial_{r}I\right)^{2}+g^{\theta\theta}\left(\partial_{\theta}I\right)^{2}+
g^{\phi\phi}\left(\partial_{\phi}I\right)^{2}\right]\nonumber\\
+A\beta\sqrt{f}\partial_{r}I \left[g^{rr}\left(\partial_{r}I\right)^{2}+g^{\theta\theta}\left(\partial_{\theta}I\right)^{2}
+g^{\phi\phi}\left(\partial_{\phi}I\right)^{2}\right]+Bm\left(1-\beta m^{2}\right)\nonumber\\
+A\frac{eA_t}{\sqrt f}\left[1-\beta m^{2} -\beta\left(g^{rr}\left(\partial_{r}I\right)^{2}+g^{\theta\theta}\left(\partial_{\theta}I\right)^{2}+
g^{\phi\phi}\left(\partial_{\phi}I\right)^{2}\right)\right]
= 0,
\label{eq3.9}
\end{eqnarray}

\begin{eqnarray}
A\left\{-(1-\beta m^2) \sqrt {g^{\theta \theta}}\partial _ {\theta} I
+ \beta \sqrt {g^{\theta \theta}}\partial _ {\theta}I\left[g^{rr}(\partial _r I)^2 + g^{\theta \theta}(\partial _ {\theta}I)^2
+ g^{\phi \phi}(\partial _ {\phi}I)^2 \right]\right.\nonumber\\
\left.-i(1-\beta m^2) \sqrt {g^{\phi \phi}}\partial _ {\phi}I+ i\beta \sqrt {g^{\phi \phi}}\partial _ {\phi}I\left[g^{rr}
(\partial _r I)^2 + g^{\theta \theta}(\partial _ {\theta}I)^2 + g^{\phi \phi}(\partial _ {\phi}I)^2\right]\right\} = 0,
\label{eq3.10}
\end{eqnarray}

\begin{eqnarray}
B\left\{-(1-\beta m^2) \sqrt {g^{\theta \theta}}\partial _ {\theta} I
+ \beta \sqrt {g^{\theta \theta}}\partial _ {\theta}I\left[g^{rr}(\partial _r I)^2 + g^{\theta \theta}(\partial _ {\theta}I)^2
+ g^{\phi \phi}(\partial _ {\phi}I)^2 \right]\right.\nonumber\\
\left.-i(1-\beta m^2) \sqrt {g^{\phi \phi}}\partial _ {\phi}I+ i\beta \sqrt {g^{\phi \phi}}\partial _ {\phi}I\left[g^{rr}
(\partial _r I)^2 + g^{\theta \theta}(\partial _ {\theta}I)^2 + g^{\phi \phi}(\partial _ {\phi}I)^2\right]\right\} = 0.
\label{eq3.11}
\end{eqnarray}

\noindent It is difficult to solve the action from the above equations. Considering properties of the dilatonic spacetime and the above equations, we carry out separation of variables on the action and get

\begin{equation}
I = -\omega t + W(r) + \Xi (\theta , \phi),
\label{eq3.12}
\end{equation}

\noindent where $\omega$ is the energy of the emitted fermion.

From eqns. (\ref{eq3.8})-(\ref{eq3.11}), it is found that eqn. (\ref{eq3.10}) and eqn. (\ref{eq3.11}) are irrelevant to $A$ and $B$ and can be reduced into the same equation. Then inserting eqn. (\ref{eq3.12}) into eqn. (\ref{eq3.10}) and eqn. (\ref{eq3.11}) and rewriting the equation yield

\begin{eqnarray}
\left(\sqrt {g^{\theta \theta}}\partial _ {\theta} \Xi +i\sqrt {g^{\phi \phi}}\partial _ {\phi}\Xi\right)\times \nonumber\\
\left[\beta g^{rr}(\partial _r W)^2 + \beta g^{\theta \theta}(\partial _ {\theta}\Xi)^2
+\beta g^{\phi \phi}(\partial _ {\phi}\Xi)^2 +\beta m^2 -1\right] =0,
\label{eq3.13}
\end{eqnarray}

\noindent which implies

\begin{eqnarray}
\sqrt {g^{\theta \theta}}\partial _ {\theta} \Xi +i\sqrt {g^{\phi \phi}}\partial _ {\phi}\Xi =0.
\label{eq3.14}
\end{eqnarray}

\noindent Thus the solution of $\Xi$ can be gotten. It is a complex function other than the trivial solution of constant. This complex function produces a contribution on the action. However, it has no contribution on the tunelling rate since the contributions of the outgoing and ingoing solutions are canceled in the calculation. From the above equation, it is easily to derive the relation

\begin{eqnarray}
g^{\theta\theta}\left(\partial_{\theta}\Xi\right)^{2}+g^{\phi\phi}\left(\partial_{\phi}\Xi\right)^{2} = 0.
\label{eq3.15}
\end{eqnarray}

\noindent Now we focus our attention on the radial action. Inserting eqn. (\ref{eq3.12}) into eqns. (\ref{eq3.8}) and (\ref{eq3.9}) and using eqn. (\ref{eq3.15}), we get

\begin{eqnarray}
B\frac{\omega}{\sqrt{f}}-B\left(1-\beta m^{2}\right)\sqrt{f}\partial_{r}W - Am\beta g^{rr}\left(\partial_{r}W\right)^{2}
+B\beta\sqrt{f} g^{rr}\left(\partial_{r}W\right)^{3}\nonumber\\
-B\frac{eA_t}{\sqrt f}\left(1-\beta m^{2} -\beta g^{rr}\left(\partial_{r}W\right)^{2}\right)+Am\left(1-\beta m^{2}\right)
= 0,
\label{eq3.16}
\end{eqnarray}

\begin{eqnarray}
-A\frac{\omega}{\sqrt{f}} -A\left(1-\beta m^{2}\right)\sqrt{f}\partial_{r}W -Bm\beta g^{rr}\left(\partial_{r}W\right)^{2}+B\beta\sqrt{f} g^{rr}\left(\partial_{r}W\right)^{3}\nonumber\\
+A\frac{eA_t}{\sqrt f}\left(1-\beta m^{2} -\beta g^{rr}\left(\partial_{r}W\right)^{2}\right)+Bm\left(1-\beta m^{2}\right)
= 0.
\label{eq3.17}
\end{eqnarray}

\noindent In the above equations, $A$ and $B$ are irrelevant to the result. Eliminating them yields

\begin{equation}
C_6\left(\partial_{r}W\right)^{6}+C_4\left(\partial_{r}W\right)^{4}+C_2\left(\partial_{r}W\right)^{2}+C_0=0,
\label{eq3.18}
\end{equation}

\noindent where

\begin{eqnarray}
C_6 &=& \beta^{2}f^{4},\nonumber\\
C_4 &=& \beta f^{3}\left(m^{2}\beta-2\right)-\beta ^2 f^2 e^2A_t^2,\nonumber\\
C_2 &=& f^2\left(1-\beta ^2 m^{4}\right) +2\beta f eA_t \left[- \omega + eA_t(1-\beta m^2)\right],\nonumber\\
C_0 &=& -m^{2}f\left(1-\beta m^{2}\right)^{2}-\left[\omega -eA_t\left(1-\beta m^{2}\right)\right]^2.
\label{eq3.19}
\end{eqnarray}

\noindent Keeping the leading order terms of $\beta$ and solving $W$ at the event horizon, we derive the imaginary part of the radial action

\begin{eqnarray}
Im W_{\pm} & = &\pm\int dr\frac{1}{f}\sqrt{m^{2}f+\left[\omega - e A_{t} (1-\beta m^2)\right]^2} \left(1+\beta m^{2}+\beta\frac{\tilde \omega^{2}}{f}- \frac{\beta eA_t \tilde\omega}{f}\right) \nonumber \\
& = & \pm \pi\frac{r_+}{\left(1 - \frac{r_-}{r_+}\right)^{\frac{1-\alpha^2}{1+\alpha^2}}} (\omega - eA_{t+}) \times\left(1 + \beta\Pi\right),
\label{eq3.20}
\end{eqnarray}

\noindent where

\begin{eqnarray}
\tilde\omega &=& \omega - eA_{t}, \nonumber\\
\Pi &=& m^2+\frac{m^2}{\omega_0}eA_{t+} + \frac{1}{2} \frac{m^2\left(\omega_0 - eA_{t+} \right)}{\omega -eA_{t+}(1- \beta m^2)} - \frac{eA_{t+}\left(\omega_0 - eA_{t+} \right)}{\left(1-\frac{r_-}{r_+}\right)^{\frac{1 - \alpha ^2}{1+\alpha ^2}}} \nonumber\\
&& + \frac{\omega_0 }{\left(1-\frac{r_-}{r_+}\right)^{\frac{1 - \alpha ^2}{1+\alpha ^2}}}\left[ 2\omega_0 +eA_{t+} - \frac{r_- (1 - \alpha ^2)\left(2\omega_0 -eA_{t+} \right)}{(r_+-r_-)(1+\alpha ^2)} - \frac{2e^2Q^2}{\omega_0 r_+} \right],
\label{eq3.21}
\end{eqnarray}

\noindent $+(-)$ denotes the outgoing(ingoing) solutions, $\omega_0 = \omega - eA_{t+}$, $A_{t} = \frac{Q}{r}$,  $A_{t+} = \frac{Q}{r_+}$ is the electromagnetic potential at the event horizon. Using the relations between $M$, $Q$ and $r_{\pm}$, it is found that $\Pi >0$. Thus the tunnelling rate of the fermion with spin up state at the event horizon is

\begin{eqnarray}
\Gamma & = & \frac{P_{(emission)}}{P_{(absorption)}} = \frac{\exp{\left(-\frac{2}{\hbar}ImI_+\right)}}{\exp{\left(-\frac{2}{\hbar}ImI_-\right)}} = \frac{exp\left( -\frac{2}{\hbar}Im W_+ - \frac{2}{\hbar}Im {\Xi}\right)}{exp\left( -\frac{2}{\hbar}Im W_- -\frac{2}{\hbar}Im {\Xi}\right)}\nonumber\\
& = & exp\left[ -4\pi\frac{r_+}{\hbar\left(1 - \frac{r_-}{r_+}\right)^{\frac{1-\alpha^2}{1+\alpha^2}}}(\omega - eA_{t+}) \times\left(1 + \beta\Pi\right)\right].
\label{eq3.22}
\end{eqnarray}

\noindent This is the Boltzmann factor with the Hawking temperature at the event horizon of the dilatonic black hole taking

\begin{eqnarray}
T = \frac{\hbar\left(1 - \frac{r_-}{r_+}\right)^{\frac{1-\alpha^2}{1+\alpha^2}}}{4\pi r_+\left(1 + \beta\Pi\right)}= T_0\left(1 -\beta\Pi\right) ,
\label{eq3.23}
\end{eqnarray}

\noindent where $T_0=\frac{\hbar}{4\pi r_+}\left(1 - \frac{r_-}{r_+}\right)^{\frac{1-\alpha^2}{1+\alpha^2}}$ is the standard Hawking temperature. It is evidently that the corrected Hawking temperature appears and is lower than the standard one. This temperature is also lower than that derived by the semi-classical tunnelling method \cite{PW}. The correction value is determined not only by the mass and charge of the black hole but also by the quantum number (energy, mass and charge) of the emitted fermion. Due to the radiation, the temperature increases. Eqn. (\ref{eq3.23}) shows that the quantum gravity correction slows down the increase of the temperature during the radiation. This correction therefore causes the radiation ceased at some particular temperature, leaving the remnant mass.

Eqn. (\ref{eq3.23}) describes the temperature of the Reissner-Nordstrom black hole when $\alpha = 0$ and that of the Schwarzschild black hole when  $\alpha = 0$ and $Q= 0$. Using an assumption that the emitted particle is massless, we estimate the remnant of the Schwarzschild black hole. The evaporation stops when $\left(M-dM\right)\left(1+\beta \omega^2\right) \simeq M$. With the observation $dM=\omega$ and $\beta = \frac{\beta_0}{M_p^2} $, we get the the remnant as $M_{R} \simeq \frac{M_p^2}{\beta_0 \omega} \geq \frac{M_p}{\beta_0}$, where we assumed that the maximal energy of the emitted particle is the Planck mass $M_p$ .

\section{The remnant in the 5-dimensional black string}

In this section, we investigate the remnant by the fermion's tunnelling from a 5-dimensional spacetime. The emitted fermion is supposed to be uncharged. Therefore, the electromagnetic effect in eqn. (\ref{eq2.9}) is not taken into account here. The 4-dimensional Schwarzschild metric is a static spherically symmetric solution to the vacuum Einstein equations. When an extra compact spatial dimension $z$ is added, the metric becomes

\begin{eqnarray}
ds^{2}=-F\left(r\right)dt^{2}+ \frac{1}{F\left(r\right)}dr^{2}+r^{2}\left(d\theta^{2}+\sin^{2}\theta d\phi^{2}\right) + dz^2,
\label{eq4.1}
\end{eqnarray}

\noindent  where $F = 1-\frac{r_h}{r}$, $r_h= 2M$ is the location of the event horizon and $M$ is proportional to the black hole mass. The metric (\ref{eq4.1}) describes a neutral uniform black string. Here we investigate a fermion tunnelling from this string.

We still only consider the spin up state. The wave function of the fermion with spin up state is now assumed as

\begin{eqnarray}
\Psi=\left(\begin{array}{c}
A\\
0\\
B\\
0
\end{array}\right)\exp\left(\frac{i}{\hbar}I\left(t,r,\theta,\phi,z\right)\right),
\label{eq4.2}
\end{eqnarray}

\noindent $A$ and $B$ are functions of $(t,r,\theta, \phi, z)$. The fermion's motion satisfies the generalized Dirac equation. Now the tetrad is different from that in the above section. It is $e_{\mu}^a = \rm{diag}\left(\sqrt{F}, 1/\sqrt{F}, r, r\sin\theta,1 \right)$. Then gamma matrices are constructed  as follows

\begin{eqnarray}
\gamma^{t} &=&\frac{1}{\sqrt{F\left(r\right)}}\left(\begin{array}{cc}
i & 0\\
0 & -i
\end{array}\right),\hspace{7mm}  \gamma^{\theta}=\sqrt{g^{\theta\theta}}\left(\begin{array}{cc}
0 & \sigma^{1}\\
\sigma^{1} & 0
\end{array}\right),\nonumber \\
\gamma^{r} &=&\sqrt{F\left(r\right)}\left(\begin{array}{cc}
0 & \sigma^{3}\\
\sigma^{3} & 0
\end{array}\right), \hspace{7mm}  \gamma^{\phi}=\sqrt{g^{\phi\phi}}\left(\begin{array}{cc}
0 & \sigma^{2}\\
\sigma^{2} & 0
\end{array}\right),\nonumber \\
\gamma^{z}&=&\left(\begin{array}{cc}
-I & 0\\
0 & I
\end{array}\right).
\label{eq4.4}
\end{eqnarray}

\noindent To apply the WKB approximation, we insert the wave function and the gamma matrices into the Dirac equation and divide by the exponential term. Multiplying by $\hbar$, the equations to leading order in $\hbar$ are obtained and decoupled into four equations

\begin{eqnarray}
-iA\frac{1}{\sqrt{F}}\partial_{t}I-B\left(1-\beta m^{2}\right)\sqrt{F}\partial_{r}I +Am\left(1-\beta m^{2}\right)+ A\left(1-\beta m^{2}\right)\partial_{z}I
\nonumber\\
+\beta \left[g^{rr}\left(\partial_{r}I\right)^{2}+g^{\theta\theta}\left(\partial_{\theta}I\right)^{2}
+g^{\phi\phi}\left(\partial_{\phi}I\right)^{2}+\left(\partial_{z}I\right)^{2}\right]\left(B\sqrt{F}\partial_{r}I-A\partial_{z}I-Am\right)
= 0,
\label{eq4.5}
\end{eqnarray}

\begin{eqnarray}
iB\frac{1}{\sqrt{F}}\partial_{t}I-A\left(1-\beta m^{2}\right)\sqrt{F}\partial_{r}I +Bm\left(1-\beta m^{2}\right)- B\left(1-\beta m^{2}\right)\partial_{z}I
\nonumber\\
+\beta \left[g^{rr}\left(\partial_{r}I\right)^{2}+g^{\theta\theta}\left(\partial_{\theta}I\right)^{2}
+g^{\phi\phi}\left(\partial_{\phi}I\right)^{2}+\left(\partial_{z}I\right)^{2}\right]\left(A\sqrt{F}\partial_{r}I+B\partial_{z}I-Bm\right)
= 0,
\label{eq4.6}
\end{eqnarray}

\begin{eqnarray}
A\left( \sqrt {g^{\theta \theta}}\partial _ {\theta} I+i\sqrt {g^{\phi \phi}}\partial _ {\phi}I\right)\times \nonumber\\
\left[g^{rr}(\partial _r I)^2 + g^{\theta \theta}(\partial _ {\theta}I)^2
+ g^{\phi \phi}(\partial _ {\phi}I)^2 + (\partial _ {z}I)^2+\beta m^2-1\right] = 0.
\label{eq4.7}
\end{eqnarray}

\begin{eqnarray}
B\left( \sqrt {g^{\theta \theta}}\partial _ {\theta} I+i\sqrt {g^{\phi \phi}}\partial _ {\phi}I\right)\times \nonumber\\
\left[g^{rr}(\partial _r I)^2 + g^{\theta \theta}(\partial _ {\theta}I)^2
+ g^{\phi \phi}(\partial _ {\phi}I)^2 + (\partial _ {z}I)^2+\beta m^2-1\right] = 0.
\label{eq4.8}
\end{eqnarray}

\noindent It is also difficult to solve the action from the above equations. Considering the property of the black string spacetime, we carry out separation of variables as

\begin{eqnarray}
I = -\omega t +W(r)+\Theta(\theta, \phi) + J z,
\label{eq4.9}
\end{eqnarray}

\noindent where $\omega$ is the energy, $J$ is a conserved momentum and describes a constant of motion corresponding to the compact dimension $z$.

We first focus our attention on the last two equations. They are irrelevant to $A$ and $B$ and can be reduced to the same equation. Inserting eqn. (\ref{eq4.9}) into them, we get  $\sqrt {g^{\theta \theta}}\partial _ {\theta} \Theta +i\sqrt {g^{\phi \phi}}\partial _ {\phi}\Theta = 0$ since the summation of factors in the square brackets in eqn. (\ref{eq4.7}) and (\ref{eq4.8}) can not be zero. Thus the solution of $\Theta$ is a complex function (other than the constant solution). The following relation,

\begin{eqnarray}
g^{\theta \theta}\left(\partial _ {\theta} \Theta\right)^2+g^{\phi \phi} \left(\partial _ {\phi}\Theta\right)^2 = 0,
\label{eq4.10}
\end{eqnarray}

\noindent is easily obtained. Return to eqns. (\ref{eq4.6}) and (\ref{eq4.7}). Inserting eqns. (\ref{eq4.9}) and (\ref{eq4.10}) into them and eliminating $A$ and $B$, we get the equation of the radial action

\begin{eqnarray}
D_6 \left(\partial _r W\right)^6+D_4 \left(\partial _r W\right)^4+D_2 \left(\partial _r W\right)^2+D_0 =0
\label{eq4.11}
\end{eqnarray}

\noindent where

\begin{eqnarray}
D_6 &=& \beta ^2F^4,\nonumber\\
D_4 &=& -2\beta x F^3 - \left(m^2 - J\right)\beta ^2F^3,\nonumber\\
D_2 &=& x^2 F^2 +2\beta x F^2\left(m^2 - J\right) - i2\beta J \omega F^{3/2},\nonumber\\
D_0 &=& -\left(m^2 - J\right)x^2F-\omega^2 +i2J\omega x\sqrt{F},\nonumber\\
x &=& 1-\beta m^2- \beta J^2.
\label{eq4.12}
\end{eqnarray}

\noindent Neglect higher order terms of $\beta$ and solve the equation (\ref{eq4.11}) at the event horizon. We only interest the imaginary part of the action because the tunnelling rate is determined by it. The imaginary part is

\begin{eqnarray}
Im W_{\pm} &=& \pm Im\int{dr\frac{\sqrt{\omega^2+\left(m^2-J^2\right)F -i2\sqrt{F}J\omega}}{F}\left[1+ \beta \left(m^2-J^2+\frac{\omega^2-2iJ\omega\sqrt{F}}{F}\right)\right]}\nonumber\\
&=&\pm \pi \omega r_h \left[1+\frac{1}{2} \beta\left(3m^2+4\omega^2-2J^2\right)\right],
\label{eq4.13}
\end{eqnarray}

\noindent where $+(-)$ are the outgoing(ingoing) solutions. Thus the tunnelling rate of the uncharged fermion across the event horizon of the 5-dimensional black string is

\begin{eqnarray}
\Gamma & = & \frac{\exp{\left(-\frac{2}{\hbar}ImI_+\right)}}{\exp{\left(-\frac{2}{\hbar}ImI_-\right)}} = \frac{exp\left( -\frac{2}{\hbar}Im W_+ - \frac{2}{\hbar}Im {\Theta}\right)}{exp\left( -\frac{2}{\hbar}Im W_- -\frac{2}{\hbar}Im {\Theta}\right)}\nonumber\\
& = & exp\left\{ -\frac{1}{\hbar}4\pi\omega r_h \left[1+\frac{1}{2} \beta\left(3m^2+4\omega^2-2J^2\right)\right]\right\},
\label{eq4.14}
\end{eqnarray}

\noindent which shows that the Hawking temperature is

\begin{eqnarray}
T & = & \frac{\hbar}{4\pi r_h \left[1+\frac{1}{2} \beta\left(3m^2+4\omega^2-2J^2\right)\right]}\nonumber\\
& = & T_0\left[1-\frac{1}{2} \beta\left(3m^2+4\omega^2-2J^2\right)\right] .
\label{eq4.15}
\end{eqnarray}

\noindent In the above equation, $T_0 = \frac{\hbar}{4\pi r_h }$ is the standard Hawking temperature. It shows that the corrected Hawking temperature is not only determined by the quantum number (energy and mass) of the emitted fermion but also affected by the effect of the extra compact dimension.

It is of interest to discuss the value of $3m^2+4\omega^2-2J^2$. When $3m^2+4\omega^2>2J^2$, it is easily found that the corrected temperature is lower than the standard Hawking temperature. This implies that the combination of the quantum gravity correction and the effect of the extra compact dimension slows down the increase of the temperature caused by the radiation. Finally, the black string should be in a stable balanced state. At this state, the remnant is left. The special case is $J =0$. In this case, the fermion's motion is limited in the 4-dimensional spacetime. Thus eqn. (\ref{eq4.15}) reduces to the temperature of the 4-dimensional Schwarzschild black hole.

When $3m^2+4\omega^2<2J^2$, the corrected temperature is higher than the standard one. It shows that the black string accelerates the evaporation and there is no remnant left. If $3m^2+4\omega^2=2J^2$, the effect of the quantum gravity correction and that of the extra dimension are canceled. Then the standard Hawking temperature appears and results in the complete evaporation. Therefore, the evaporation of the black string is affected by the quantum gravity correction and the effect of the extra compact dimension.

\section{Conclusion}

In this paper, based on the modified fundamental commutation relation, we modified the HUP and investigated the fermions' tunnelling across the horizons of the 4-dimensional charged dilatonic black hole and the 5-dimensional neutral black string. The corrected Hawking temperatures were gotten. The remnants were discussed by the temperatures. For the dilatonic black hole, the correction is determined not only by the mass and charge of the black hole but also by the quantum number (mass, charge and energy) of the emitted fermion. The interesting point is that the quantum gravity correction slows down the increase of the Hawking temperature. It is natural to lead to the remnant left in the evaporation. For the black string, the temperature is affected by the quantum number (mass and energy) of the emitted fermion and the effect of extra compact dimension. The existence of the remnant is determined by the combined effect of the quantum gravity correction and the compact dimension. In \cite{NIC4}, noncommutative black holes was discussed. In \cite{RMS,SVZR}, the quantum tunnelling radiation were researched beyond the semiclassical approximation. The corrected Hawking temperatures were derived and also lower than the standard semiclassical Hawking temperatures.

\acknowledgments

This work is supported in part by the NSFC (Grant Nos. 11205125, 11005016, 11005086, 11175039 and 11375121), the Innovative Research Team in College of Sichuan Province (Grant No. 13TD0003) and  SiChuan Province Science Foundation for Youths (Grant No. 2012JQ0039).


\bigskip

\begin{thebibliography}{99}

\small


\bibitem{SWH}
S.W. Hawking, \emph{Particle creation by black holes}, \emph{Math. Phys.} \textbf{43} (1975) 199.

\bibitem{PW}
M.K. Parikh and F. Wilczek, \emph{Hawking radiation as tunneling}, \emph{Phys. Rev. Lett.} \textbf{85} (2000) 5042 [arXiv:hep-th/9907001].

\bibitem{ZZ}
J.Y. Zhang and Z. Zhao, \emph{Hawking radiation of charged particles via tunneling from the Reissner-Nordstrom black hole}, \emph{JHEP} \textbf{10} (2005) 055.

J.Y. Zhang and Z. Zhao, \emph{Charged particles' tunnelling from the Kerr-Newman black hole}, \emph{Phys. Lett.} \textbf{B 638} (2006) 110 [arXiv:0512153[gr-qc]].

\bibitem{JWC}
Q.Q. Jiang, S.Q. Wu and X. Cai, \emph{Hawking radiation as tunneling from the Kerr and Kerr-Newman black holes}, \emph{Phys. Rev.} \textbf{D 73} (2006) 064003; Erratum-ibid. \textbf{D 73} (2006) 069902 [arXiv:0512351[hep-th]].

Q.Q. Jiang and S.Q. Wu, \emph{Hawking radiation of charged particles as tunneling from Reissner-Nordstrom-de Sitter black holes with a global monopole}, \emph{Phys. Lett.} \textbf{B 635} (2006) 151 [arXiv:hep-th/0511123].

\bibitem{ANVZ}
M. Angheben, M. Nadalini, L. Vanzo and S. Zerbini, \emph{Hawking radiation as tunneling for extremal and rotating black holes}, \emph{JHEP} \textbf{05} (2005) 014 [arXiv:hep-th/0503081].

\bibitem{KM1}
R. Kerner and R.B. Mann, \emph{Tunnelling, temperature and Taub-NUT black holes}, \emph{Phys. Rev.} \textbf{D 73} (2006) 104010 [arXiv:gr-qc/0603019].

\bibitem{KM2}
R. Kerner and R.B. Mann, \emph{Fermions tunnelling from black holes}, \emph{Class. Quant. Grav.} \textbf{25} (2008) 095014 [arXiv:0710.0612 [hep-th]].

R. Kerner and R.B. Mann, \emph{Charged fermions tunnelling from Kerr-Newman black holes}, \emph{Phys. Lett.} \textbf{B 665} (2008) 277 [arXiv:0803.2246 [hep-th]].

\bibitem{AAS}
S.P. Robinson and F. Wilczek, \emph{Relationship between Hawking radiation and gravitational anomalies}, \emph{Phys. Rev. Lett.} \textbf{95} (2005) 011303.

S. Iso, H. Umetsu and F. Wilczek, \emph{Hawking radiation from charged black holes via gauge and gravitational anomalies}, \emph{Phys. Rev. Lett.} \textbf{96} (2006) 151302.

S. Hemming and E. Keski-Vakkuri, \emph{The spectrum of strings on BTZ black holes and spectral flow in the SL(2,R) WZW model}, \emph{Phys. Rev.} \textbf{D 64} (2001) 044006 [arXiv: hep-th/0110252].

E.C. Vagenas, \emph{Are extremal 2D black holes really frozen?}, \emph{Phys. Lett.} \textbf{B 503} (2001) 399 [arXiv:hep-th/0012134].

A.J.M. Medved, \emph{Radiation via tunneling from a de Sitter cosmological horizon}, \emph{Phys. Rev.} \textbf{D 66} (2002) 124009 [arXiv:hep-th/0207247].

M. Arzano, A.J.M. Medved and E.C. Vagenas, \emph{Hawking radiation as tunneling through the quantum horizon}, \emph{JHEP} \textbf{0509} (2005) 037 [hep-th/0505266]

P. Mitra, \emph{Hawking temperature from tunnelling formalism}, \emph{Phys. Lett.} \textbf{B 648} (2007) 240 [arXiv:hep-th/0611265].

S.Q. Wu, Q.Q. Jiang, \emph{Remarks on Hawking radiation as tunneling from the BTZ black holes}, \emph{JHEP} \textbf{0603} (2006) 079 [arXiv:hep-th/0602033].

E.T. Akhmedov, V. Akhmedova and D. Singleton, \emph{Hawking temperature in the tunneling picture}, \emph{Phys. Lett.} \textbf{B 642} (2006) 124 [arXiv:hep-th/0608098].

B.D. Chowdhury, \emph{Problems with tunneling of thin shells from black holes}, \emph{Pramana} \textbf{70} (2008) 593 [arXiv:hep-th/0605197].

B. Chatterjee, A. Ghosh and P. Mitra, \emph{Tunnelling from black holes and tunnelling into white holes}, \emph{Phys. Lett.} \textbf{B 661} (2008) 307 [arXiv:0704.1746 [hep-th]].

R. Banerjee and B.R. Majhi, \emph{Quantum tunneling and back reaction}, \emph{Phys. Lett.} \textbf{B 662} (2008) 62 [arXiv:0801.0200[hep-th]].

\bibitem{PKT}
P.K. Townsend, \emph{Small-scale structure of spacetime as the origin of the gravitational constant}, \emph{Phys. Rev.} \textbf{D 15} (1977) 2795.

\bibitem{ACV}
D. Amati, M. Ciafaloni and G. Veneziano, \emph{Can spacetime be probed below the string size?} \emph{Phys. Lett.} \textbf{B 216} (1989) 41.

\bibitem{KPP}
K. Konishi, G. Paffuti and P. Provero, \emph{Minimum physical length and the generalized uncertainty principle in string theory}, \emph{Phys. Lett.} \textbf{B 234} (1990) 276.

\bibitem{LJG}
L.J. Garay, \emph{Quantum gravity and minimum length}, \emph{Int. J. Mod. Phys.} \textbf{A 10} (1995) 145 [arXiv:gr-qc/9403008].

\bibitem{GAC}
G. Amelino-Camelia, \emph{Relativity in space-times with short-distance structure governed by an observer-independent (Planckian) length scale}, \emph{Int. J. Mod. Phys.} \textbf{D 11} (2002) 35 [arXiv:gr-qc/0012051].

\bibitem{NIC2}
M.~Sprenger, P.~Nicolini and M.~Bleicher, \emph{Physics on Smallest Scales - An Introduction to Minimal Length Phenomenology}, \emph{Eur. J. Phys.}  {\bf C 33}, 853 (2012)  [arXiv:1202.1500 [physics.ed-ph]].

\bibitem{FS}
F. Scardigli, \emph{Generalized uncertainty principle in quantum gravity from micro-black hole gedanken experiment}, \emph{Phys. Lett.} \textbf{B 452} (1999) 39 [arXiv:hep-th/9904025].

\bibitem{KMM}
A. Kempf, G. Mangano, R.B. Mann, \emph{Hilbert space representation of the minimal length uncertainty relation}, \emph{Phys. Rev.} \textbf{D 52} (1995) 1108 [arXiv:hep-th/9412167].

\bibitem{AK}
A. Kempf, \emph{Nonpointlike particles in harmonic oscillators}, \emph{J. Phys.} \textbf{A 30} (1997) 2093
[arXiv:hep-th/9604045].

\bibitem{FB}
F. Brau, \emph{Minimal length uncertainty relation and Hydrogen atom}, \emph{J. Phys.} \textbf{A 32} (1999) 7691
[arXiv:quant-ph/9905033].

\bibitem{ADV}
A.F. Ali, S. Das and E.C. Vagenas, \emph{Discreteness of space from the generalized uncertainty
principle}, \emph{Phys. Lett.} \textbf{B 678} (2009) 497 [arXiv:0906.5396[hep-th]].

\bibitem{DV}
S. Das and E.C. Vagenas, \emph{Universality of quantum gravity corrections}, \emph{Phys. Rev. Lett.} \textbf{101} (2008) 221301 [arXiv:0810.5333[hep-th]].

\bibitem{NIC3}
A.~Smailagic and E.~Spallucci, \emph{Lorentz invariance, unitarity in UV-finite of QFT on noncommutative spacetime},  \emph{J. Phys.}  {\bf A 37}, 1 (2004) [hep-th/0406174].

\bibitem{AFA}
A.F. Ali, \emph{No existence of black holes at LHC due to minimal length in quantum gravity}, \emph{JHEP} \textbf{1209} (2012) 067 [arXiv:1208.6584[hep-th]].

J.~Mureika, P.~Nicolini and E.~Spallucci, \emph{Could any black holes be produced at the LHC?}, \emph{Phys. Rev.} {\bf D 85} (2012) 106007 [arXiv:1111.5830 [hep-ph]].

\bibitem{XW}
L. Xiang, X.Q. Wen, \emph{Black hole thermodynamics with generalized uncertainty principle}, \emph{JHEP} \textbf{0910} (2009) 046 [arXiv:0901.0603[qr-qc]].

\bibitem{BJM}
A. Bina, S. Jalalzadeh, A. Moslehi, \emph{Quantum black hole in the generalized uncertainty principle framework}, \emph{Phys. Rev.} \textbf{D 81} (2010) 023528 [arXiv:1001.0861[qr-qc]].

\bibitem{KP}
Y.M. Kim, Y.J. Park, \emph{Entropy of the Schwarzschild black hole to all orders in the Planck length}, \emph{Phys. Lett.} \textbf{B 655} (2007) 172 [arXiv:0707.2128[qr-qc]].

\bibitem{ZDM}
K. Zeynali, F. Darabi, H. Motavalli, \emph{Black hole thermodynamics and modified GUP consistent with doubly special relativity}, \emph{Mod. Phys. Lett.} \textbf{A 27} (2012) 1250227 [arXiv:1206.5121[qr-qc]].

\bibitem{ACS}
R.J. Adler, P. Chen and D.I. Santiago, \emph{The generalized uncertainty principle and black hole remnants}, \emph{Gen. Rel. Grav.} \textbf{33} (2001) 2101 [arXiv:gr-qc/0106080].

\bibitem{SGC}
F. Scardigli, C. Gruber, P. Chen, \emph{Black hole remnants in the early universe}, \emph{Phys. Rev.} \textbf{D 83} (2011) 063507 [arXiv:1009.0882[qr-qc]].

\bibitem{BG}
R. Banerjee, S. Ghosh, \emph{Generalised uncertainty principle, remnant mass and singularity problem in black hole thermodynamics}, \emph{Phys. Lett.} \textbf{B 688} (2010) 224 [arXiv:1002.2302[gr-qc]].

\bibitem{LX}
L. Xiang, \emph{A note on the black hole remnant}, \emph{Phys. Lett.} \textbf{B 647} (2007) 207. [arXiv:gr-qc/0611028].

\bibitem{NS}
K. Nozari and S. Saghafi, \emph{Natural cutoffs and quantum tunneling from black hole horizon},
\emph{JHEP} \textbf{11} (2012) 005 [arXiv:1206.5621[hep-th]].

\bibitem{NIC1}
P.~Nicolini, \emph{Nonlocal and generalized uncertainty principle black holes}, [arXiv:1202.2102 [hep-th]].

M.~Isi, J.~Mureika and P.~Nicolini, \emph{Self-Completeness and the Generalized Uncertainty Principle}, \emph{JHEP} {\bf 1311} (2013) 139 [arXiv:1310.8153 [hep-th]].

\bibitem{NM}
K. Nozari and S.H. Mehdipour, \emph{Parikh-Wilczek tunneling from noncommutative higher dimensional black holes}, \emph{JHEP} \textbf{0903} (2009) 061 [arXiv:0902.1945[hep-th]].

\bibitem{NK}
K. Nozari and M. Karami, \emph{Minimal length and generalized Dirac equation}, \emph{Mod. Phys. Lett.} \textbf{A 20} (2005) 3095 [arXiv:hep-th/0507028].

\bibitem{WG}
W. Greiner, \emph{Relativistic Quantum Mechanics: Wave Equation}, \emph{Springer-Verlag} 2000.

\bibitem{HBH}
S. Hossenfelder, M. Bleicher, S. Hofmann, J. Ruppert, S. Scherer and H. Stocker, \emph{Signatures in the Planck regime}, \emph{Phys. Lett.} \textbf{B 575} (2003) 85 [arXiv:hep-th/0305262].

\bibitem{OS}
T.L.A. Oakes, R.O. Francisco, J.C. Fabris, J.A. Nogueira, \emph{Ground State of the Hydrogen Atom via Dirac Equation in a Minimal Length Scenario}, \emph{Eur. Phys. J.} \textbf{C 73} (2013) 2495.

A. Shokrollahi, \emph{Free motion of a Dirac particle with a minimum uncertainty in position}, \emph{Reports on Mathematical Physics} \textbf{70} (2012) 1.

\bibitem{MK}
M. Kober, \emph{Gauge theories under incorporation of a generalized uncertainty principle}, \emph{Phys. Rev.} \textbf{D 82} (2010) 085017.

\bibitem{GHSHH}
D. Garfinkle, G.T. Horowitz and A. Strominger, \emph{Charged black holes in string theory}, \emph{Phys. Rev.} \textbf{D 43} (1991) 3140.

J. Horne and G.T. Horowitz, \emph{Rotating dilaton black holes}, \emph{Phys. Rev.} \textbf{D 46} (1992) 1340.

\bibitem{NIC4}
P.~Nicolini, A.~Smailagic and E.~Spallucci, \emph{Noncommutative geometry inspired Schwarzschild black hole}, \emph{Phys. Lett.} {\bf B 632} (2006) 547 [gr-qc/0510112].

P.~Nicolini, \emph{Noncommutative black holes, the final appeal to quantum gravity: A review}, \emph{Int. J. Mod. Phys.}{\bf A 24} (2009) 1229 [arXiv:0807.1939 [hep-th]].

\bibitem{RMS}
R. Banerjee and B.R. Majhi, \emph{Quantum tunneling beyond semiclassical approximation}, \emph{JHEP} \textbf{0806} (2008) 095 [arXiv:0805.2220[hep-th]].

B.R. Majhi,  \emph{Fermion tunneling beyond semiclassical approximation}, \emph{Phys. Rev.} \textbf{D 79} (2009) 044005 [arXiv:0809.1508[hep-th]].

\bibitem{SVZR}
D. Singleton, E.C. Vagenas, T. Zhu and J.R. Ren, \emph{Insights and possible resolution to the information loss paradox via the tunneling picture}, \emph{JHEP} \textbf{1008} (2010) 089 [arXiv:1005.3778 [gr-qc]].



  
  
  
\end{thebibliography}
\end{document}